\documentclass{article}

\usepackage{arxiv}

\usepackage[utf8]{inputenc} 
\usepackage[T1]{fontenc}    
\usepackage{hyperref}       
\usepackage{url}            
\usepackage{booktabs}       
\usepackage{amsfonts}       
\usepackage{nicefrac}       
\usepackage{microtype}      
\usepackage{lipsum}		
\usepackage{graphicx}
\usepackage[square,numbers]{natbib}
\usepackage{doi}
\usepackage{amsmath}

\title{Real-time tracking of COVID-19 and coronavirus research updates through text mining}

\author{ \href{https://orcid.org/0000-0001-9583-5139}{\includegraphics[scale=0.06]{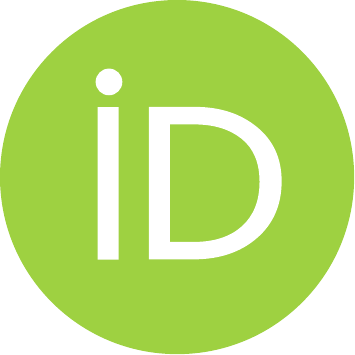}\hspace{1mm}Yutong Jin}, {\hspace{2mm}Jie Li}, {\hspace{2mm}Xinyu Wang}, {\hspace{2mm}Peiyao Li}, {\hspace{2mm}Jinjiang Guo}, {\hspace{2mm}Junfeng Wu},{\hspace{2mm}Dawei Leng}, \href{https://orcid.org/0000-0001-9321-8348} {\hspace{2mm}\includegraphics[scale=0.06]{orcid.pdf} Lurong Pan*}\\
 \emph{AIDD Group}\\
 Global Health Drug Discovery Institute, Beijing, China \\
 \texttt{yutong.jin@ghddi.org}  and \texttt{lurong.pan@ghddi.org}  \\}

\date{}

\renewcommand{\headeright}{}
\renewcommand{\undertitle}{}

\hypersetup{
pdftitle={NLPcov},
pdfsubject={cs.DS},
pdfauthor={Yutong Jin and Lurong Pan},
pdfkeywords={Natural Language Processing, Text Mining,  Clustering, COVID-19},
}

\begin{document}
\maketitle

\begin{abstract}
The novel coronavirus (SARS-CoV-2) which causes COVID-19 is an ongoing pandemic. There are ongoing studies with up to hundreds of publications uploaded to databases daily. We are exploring the use-case of artificial intelligence and natural language processing in order to efficiently sort through these publications. We demonstrate that clinical trial information, preclinical studies, and a general topic model can be used as text mining data intelligence tools for scientists all over the world to use as a resource for their own research. To evaluate our method, several metrics are used to measure the information extraction and clustering results. In addition, we demonstrate that our workflow not only have a use-case for COVID-19, but for other disease areas as well. Overall, our system aims to allow scientists to more efficiently research coronavirus. Our automatically updating modules are available on our information portal at https://ghddi-ailab.github.io/Targeting2019-nCoV/ for public viewing. 
\end{abstract}


\section{Introduction}
The COVID-19 pandemic is an ongoing pandemic caused by the novel coronavirus (SARS-CoV-2)\cite{cov19wiki}. The symptoms are highly variable, and the virus, which spreads through the air and contaminated surfaces, is highly contagious. As of January 2021, there has yet to be a small molecule drug that is specific and effective for COVID-19. During the pandemic, countries around the world made efforts to overcome the difficulties, further reflecting the importance of unity and cooperation and resource sharing. We are continuously exploring the value chain provided by artificial intelligence (AI) in the drug discovery process. In terms of pathological mechanisms, AI natural language processing (NLP) technology can replace manual curation of data and efficiently collect and sort data from global databases. 

Data mining is a process in which algorithms convert raw data into useful structured data. This technique is then integrated with NLP algorithms to analyze and organize the collected information from areas such as a disease field either through rule-based text mining or a model-based tool. In our effort, we have launched the GHDDI Targeting COVID-19 platform \cite{targetingcov19}. Since its launch on January 29, the platform has been continuously updated and maintained with new functions and modules continuously added. Several of these functional modules include NLP data mining module for SARS-CoV-2 small molecule drug in vitro experimental data that updates new experimental information daily, an automated NLP COVID-19 clinical trial module allowing up-to-date summarization of clinical trial data, and an NLP-based scientific literature recommendation module. Overall, we present the details behind these three modules to support real-time scientific intelligence of COVID-19.

\section{Methods}
\label{sec:methods}
In this section, we briefly introduce the three modules and how they were built using different databases. All of the NLP systems were built using a standard Python 3.6 environment from Anaconda and associated packages mentioned below. Our system's backend and database is hosted on a Ubuntu 18.04 server using the same environment.

\subsection{Data Aggregation and Preprocessing}
\label{sec:mdap}
The data was aggregated using a variety of sources. Through automated download scripts and given Application Programming Interfaces (API), abstract data was downloaded from: PubMed, preprint sources, and dimensions.ai \cite{dimensions2020} using a string query "SARS-Cov-2 OR COVID-19 OR novel coronavirus". These data sources were compiled together and the string data was cleaned using simple Python scripts such as lower-casing all words and removing noise data such as spaces or tabs. Duplicate data was then removed through a sequence of steps by dropping DOI, title, and abstract strings, respectively. This aggregated dataset will be used for subsequent NLP workflows and models.

\subsection{Data Dictionaries}
\label{sec:mdd}
There were several dictionaries that were compiled and utilized for information filtering and information extraction. First, a dictionary of all drug names was compiled using DrugBank drug names and aliases \cite{drugbank},  FDA drug list \cite{fdalist}, and ChEMBL \cite{chembl2015}. All of these drug names were compiled into one list, and string length was computed to filter out outliers. Overall, the final list consisted of unigram, bigrams, and trigrams; it also included drug names with a string length between 5 and 75 characters.

The second dictionary involved a filter list to clean out unwanted items from the drug dictionary. In the DrugBank database, several entries that do not necessarily represent drug names can be found such as large biologic molecules and antigens. Likewise, in a similar Kaggle competition \cite{kaggle2020}, a list of filtered items were compiled, and this list was aggregated and used to filter out unwanted terms in the final drug dictionary used in subsequent workflows.

\subsection{Preclinical Data Extraction}
\label{sec:mpde}
This step uses the aggregated dataset mentioned in the previous section. The dataset is then sent through a filter of keywords [EC50, IC50, CC50] to get a subset of only abstracts mentioning these keywords. Then, each abstract from the subset is matched with a corpus of known drug names and sentences are extracted. If sentences contain both a key word and a drug name, the sentence will be searched for a numerical value or descriptive phrase describing the relationship in that sentence. This is done using either regex (Rule 1) or a Spacy noun chunk model (Rule 2). Using regex, the system extracts the numerical value closest in word distance to the keyword or using a rule-based logic. Similarily, the Spacy noun chunk model extracts the noun chunk describing the keyword. The spacy model is an open-source English language model. Several features include POS tagging, noun chunk extraction, and grammar NER. The noun chunks are a descriptive phase that has significant relations to a keyword. This extracted values list is then extracted and mapped onto the drug name with a direct correlation. Additionally, if sentences mentioning the drug and keyword are different, then the system tries to extract a value similar to above but prints the drug name and experimental assay value relationship as an indirect correlation. These results are all tabulated and updated to the website. 

\begin{itemize}
	\item Rule 1: Using a regex query, all numbers are extracted. The closest numerical token to the experimental keyword is mapped to the closest drug name.
	\item Rule 2: Using a Spacy model, all noun chunks are extracted from the sentence. The noun chunk closest in distance to that of the experimental keyword is identified and mapped. 
\end{itemize}

Using these two rules, all data following this logic can be extracted and mapped. Because this text mining procedure is done using a list of known drugs, several metrics are used to validate this workflow. We have evaluated the text mining results based on a similar text mining study\cite{medex}. In that study, 25 unique text items (notes) were randomly sampled and manually reviewed as a gold standard. Overall, we evaluated Precision, Recall, and F-measure in the preclinical data mining results by randomly sampling 25 papers by DOI. 

\begin{equation}
\label{eq:precision}
P = N_{correct}/ N_{total}
\end{equation}

\begin{equation}
\label{eq:precision}
R = N_{correct}/ N_{total\ possible}
\end{equation}

The above equations are derived from calculating precision and recall in an information extraction context\cite{ting2010}. 

\subsection{NLP Topic Model Recommendation Engine}
\label{sec:mntmre}
The aggregated dataset built in Section 2.1 is utilized in this step. Figure 1 shows the monthly amount of articles uploaded onto our database. As a result of this large number, it was important to split the articles into different categories and recommend them by topic. After preprocessing, the data is then checked and tagged for up to trigrams. Additionally, data lemmatization is used and stop words are removed; a bag-of-words is subsequently created for each abstract. A Latent Dirichlet Allocation (LDA) algorithm is used to build the topic model. This is an unsupervised machine learning model that measures the distribution of words and attempts to cluster this distribution into a specified number of hidden distributions. The word distributions per abstract determine which topic or hidden distribution that abstract best fits into. The bag-of-words object is then sent to the Gensim LDA API \cite{lda2020} for model training, and subsequent Python pickle objects and metadata were used for daily updates. 

A gridsearch optimization of this topic model was performed by maxmizing the coherence score, and the best scoring model was used for the final output where each topic was hand-labeled. In this dataset, the best scoring model was a 30-topic model which was used for the final output. After this model was trained, inferencing was performed on the original dataset, and then each abstract was assigned a topic. This result was recorded, and a data-driven filter was used to filter out topics that did not meet the amount papers required to form a topic. Then for each topic, the top papers are ranked and sorted by the model’s output weight, being the gamma value in the Gensim model, and the top 10 papers are output into a final tabulated format. This can then be done for new data which can be automatically updated in the future for this module.

\subsection{Clinical Trials Text Mining}
\label{sec:mctdm}
In the clinical trials module, the open-access Figshare data shared by dimensions.ai was used \cite{dimensions2020}. As of January 2021, there were 7000 clinical trials records around the world. Clinical trials contain different phase human experiments relating to drugs or biologics, including vaccines. The data is first preprocessed similar to the methods mentioned above. The data is then tagged for unigrams, bigrams, and trigrams similar to the NLP topic model. Afterwards, the information extraction process clusters the clinical trials into one of these three types. 

\begin{itemize}
	\item Using the known drugs dictionary mentioned in the preclinical information extraction section, all small molecule drug names are extracted from the clinical trial descriptive phrases. This list is filtered and extracted samples containing animal, food, and other non-small molecule drug words are removed.
	\item Given the keyword “vaccine” and all its derivatives, the database was searched for these keywords and a list of vaccine clinical trials was compiled and output. This list is filtered and trials containing words in the blacklist are removed.
	\item Given several keywords relating to biological products such as plasma, antibody, stem cell, and all of their derivative words, a list of biological products was compiled and output. This list is filtered and trials containing words in the blacklist are removed.
\end{itemize}

Using these three rules, all information pertaining to drugs, biologicals, and vaccines were extracted from the tabulated data. The data was visualized in our information portal \cite{targetingcov19} together with word clouds for biological drug and vaccine trials as a validation.

\begin{figure}
	\centering
	\includegraphics[scale=0.666]{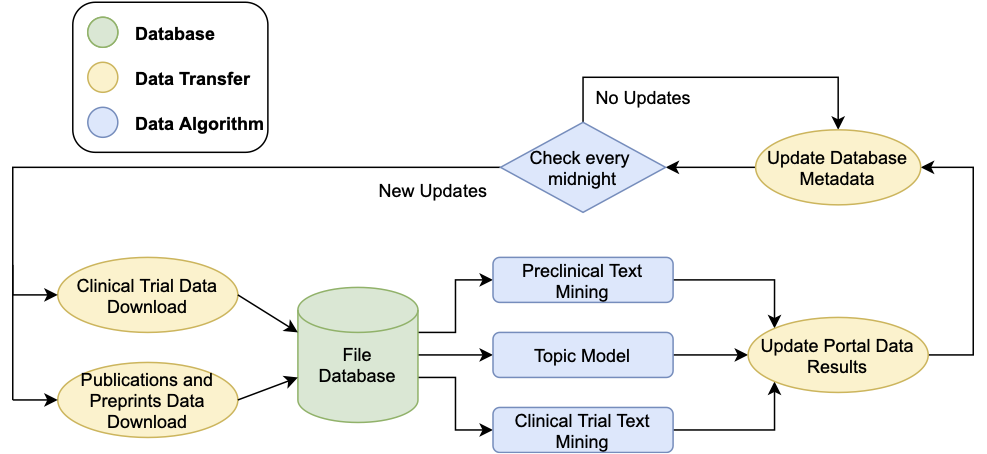}
	\caption{Workflow of real-time system to update modules}
	\label{fig:Picture0}
\end{figure}

\subsection{Real-time Updates}
\label{sec:mctdm}
All of these modules are supported by real-time daily updates provided by a server setup to update automatically. This system, as shown in Figure 1, provides daily incremental updates of clinical trial data and research articles using open APIs provided by PubMed, Figshare, and other sources. This data is then stored in our database which is updated daily. After updating the database, we use metadata to track changes in the clinical trial and research article databases. The clinical trial script is run automatically every day and completes the data processing if there is a new update. Likewise, new articles recently updated to our database is preprocessed and then run with the preclinical NLP processing workflow, and updates are appended to a master list that is updated onto our portal. Finally, the entire abstracts database is preprocessed then run with the topic model; afterwards, the top 10 titles are uploaded per topic to the recommendation page. This model is retrained and updated monthly as new data becomes available.

\section{Results}
\label{sec:results}
The results of our modules are presented below. Full results can be found on the Targeting COVID-19 GitHub portal \cite{targetingcov19}. This section gives an in-depth description of the results that were published to the website.

\begin{figure}
	\centering
	\includegraphics[scale=1]{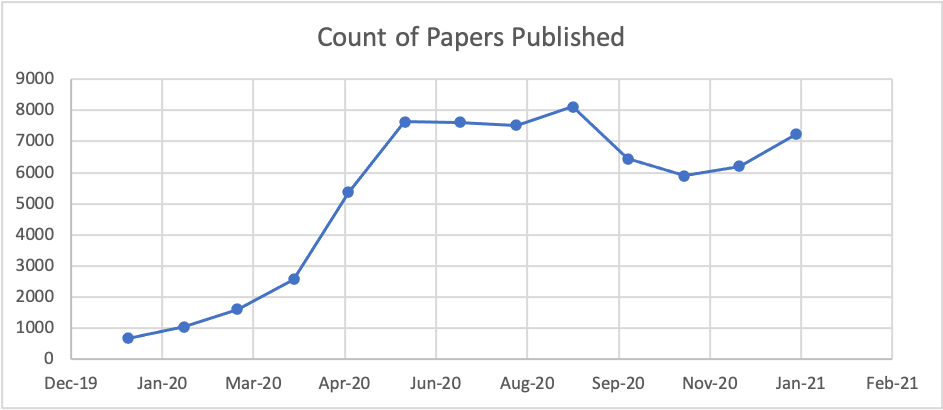}
	\caption{Count of papers published every month starting from January 2020.}
	\label{fig:Picture1}
\end{figure}

Figure 2 visualizes the number of papers uploaded to the databases by month. Due to the number of papers published exponentially increasing in March 2020, it became impossible to track all experimental and clinical results published to a journal or uploaded onto a preprint service. Therefore, we used this data to automatically data mine and extract valuable information that may be of use to scientists of different fields all around the world.

\subsection{Small Molecule Drug Text Mining}
For small molecule drugs, the compiled drug dictionary is used and the matches are tabulated in the following tables. Table 1 shows the top 10 most common drugs found through information extraction of clinical trial records where there are currently over 1100 clinical trials for small molecule drugs, while Table 2 shows several of the best experimental results of small molecule drugs extracted from preclinical studies literature text. It is noted that the units are extracted from the sentence of the experimental value; standard units are typically given as a molar concentration such as micromolar or nanomolar units.

\begin{table}
\caption{Top 20 known small molecule drugs undergoing COVID-19 clinical trials.}
\centering
\begin{tabular}{ll}
\toprule
\textbf{Treatment}                  & \textbf{Count} \\
\midrule
Hydroxychloroquine & 153 \\
Ritonavir & 65 \\
Lopinavir & 61 \\
Azithromycin & 60 \\
Tocilizumab & 55 \\
Ivermectin & 51 \\
Favipiravir & 38 \\
Remdesivir & 33 \\
Chloroquine & 32 \\
Colchicine & 24 \\
Dexamethasone & 23 \\
Methylprednisolone & 23 \\
Enoxaparin & 22 \\
Nitazoxanide & 20 \\
Ruxolitinib & 19 \\
Anakinra & 15 \\
Angiotensin & 15 \\
Heparin & 15 \\
Baricitinib & 14 \\
Interferon beta & 14 \\
\bottomrule
\end{tabular}
\label{tab:cttable}
\end{table}

\begin{table}
\caption{Five small molecule drugs with in vitro assay results.}
\centering
\begin{tabular}{llll}
\toprule
\textbf{Drug name}    & \textbf{Assay} & \textbf{Value}  & \textbf{Units (uM)} \\
\midrule
Nafamostat   & IC50  & 0.0022 & micro \\
Azithromycin & EC50  & 0.008  & um    \\
Pralatrexate & EC50  & 0.008  & um    \\
Adenosine    & EC50  & 0.01   & um    \\
Remdesivir   & EC50  & 0.01   & um    \\
\bottomrule
\end{tabular}
\label{tab:pctable}
\end{table}

\subsection{Text Mining Examples from Unstructured Abstract Text}
Using the rules previously described in the Methods section,  Figure 3 shows several examples of direct correlation sentences of nafamostat, which is also shown in Table \ref{tab:pctable}, labelled with an experimental value along with the experiment type from three different article abstracts\cite{naf1} \cite{naf2} \cite{naf3}. Nafamostat is a small molecule drug which had the best experimental value out of all of the extracted data samples. These sentences were taken directly from the preprint or published abstracts aggregated in our database, and Figure 3 visualizes what the rule-based search engine looked for in each abstract. It is noted that several other drugs are also labeled in Figure 3 for visualization purposes, but these drugs are not described in further detail. 

\begin{figure}
	\centering
	\includegraphics[scale=0.5]{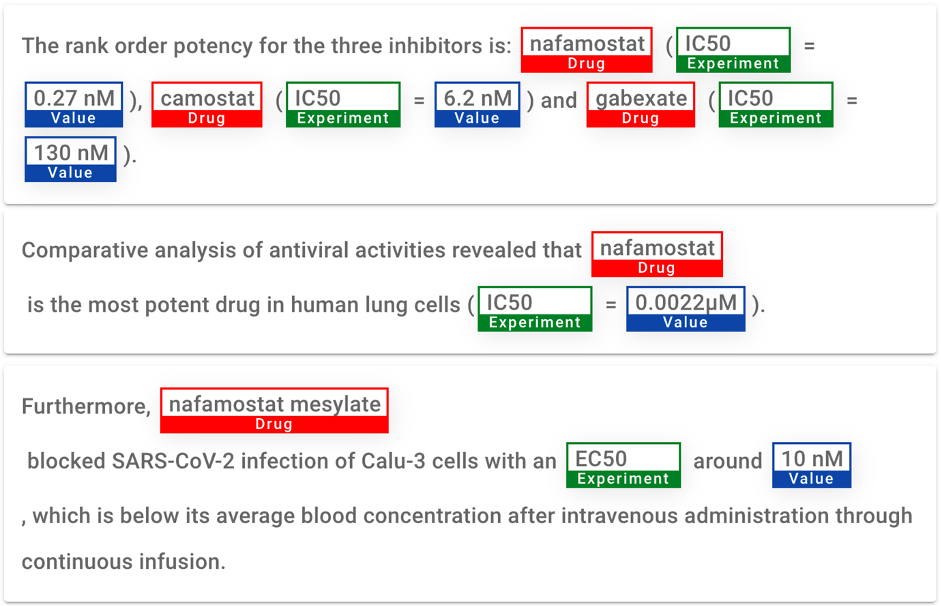}
	\caption{Several sentences from different abstracts containing the drug “nafamostat”. All drugs were labeled in red, experiments in green, and numerical values in blue.}
	\label{fig:Picture2}
\end{figure}

\subsection{Topic Model Examples}
Table 3 prints 5 of the topics taken from the LDA topic model along with their manually assigned label. The top topics were taken from a grid-search optimized number of topics while maximizing the coherence score using a corpus with over 20,000 abstracts. The best topic model was found to contain 30 topics. Among these topics, there were several that had minimal samples for that cluster. These topics were removed from the final presentation using a data driven approach. Several paper titles for each topic are shown in Figure 4 justifying the manual label attached to each LDA topic. 

\begin{table}
	\caption{Topic keywords for five select topics in our optimized topic model.}
	\centering
	\begin{tabular}{ccccc}
		\toprule
		AI        & Mental Health & Disease Analysis & Genetics  & PPE           \\
		\midrule
		covid     & covid         & covid            & sars\_cov & mask          \\
		ct        & health        & risk             & protein   & use           \\
		score     & mental        & age              & ace       & air           \\
		use       & pandemic      & high             & viral     & respirator    \\
		image     & anxiety       & population       & virus     & particle      \\
		diagnosis & participant   & mortality        & human     & surface       \\
		pneumonia & report        & factor           & cell      & wear          \\
		feature   & study         & infection        & host      & environmental \\
		lung      & survey        & disease          & analysis  & device        \\
		base      & psychological & increase         & genome    & transmission        \\
		\bottomrule
	\end{tabular}
	\label{tab:tmtable1}
\end{table}

\begin{figure}
	\centering
	\caption{Select Titles of the five topics in our optimized topic model in Table 3. Some of the title names were truncated because of the large string size.}
	\includegraphics[scale=.55]{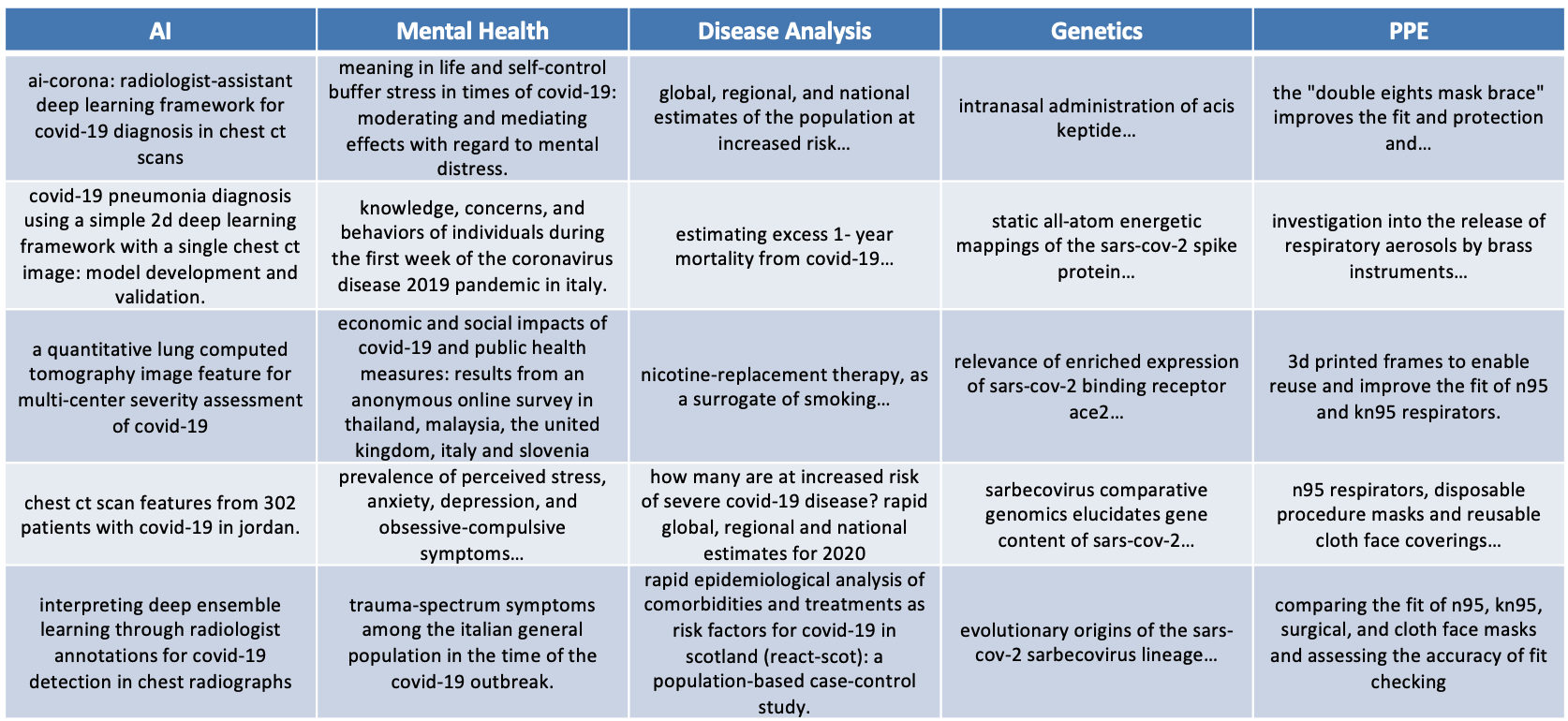}
	\label{fig:table4}
\end{figure}

\section{Discussion}
\label{sec:discuss}
We have developed several automatic modules from the openly available data. The full data results are publicly available at our website \textbf{COVID-19: GHDDI Info Sharing Portal}: \url{https://ghddi-ailab.github.io/Targeting2019-nCoV}

\subsection{Evaluation of Text Mining Results}
We evaluated a random subset of article abstracts with a gold standard that was manually read and labeled for the same information extraction task. The manual gold standard results are then evaluated with the results in Table 4.

\begin{table}
\caption{Results of data extraction system on a random subset of abstracts.}
\centering
\begin{tabular}{ll}
\toprule
\textbf{Metric} & \textbf{Value} \\
\midrule
Precision & 0.808 \\
Recall & 0.689 \\
F1 Score & 0.743 \\
\bottomrule
\end{tabular}
\label{tab:dm_pr}
\end{table}

It can be seen that the precision of the system was assessed to be around 0.8 showing that our system can indeed extract most of the drug names and experimental values correctly. Therefore, this validates that our system can be used as a recommendation for users to follow-up on these articles. After a manual review of the articles, it was found that many of the drug names that could not be extracted were not found in our drug dictionary that we had compiled. Additionally, wrong experimental values and wrong drug name mappings were attributed to the fact that the rule-based system cannot robustly handle some content such as when multiple numerical values appear at multiple locations in a sentence. More fine-tuning of this system’s rules is needed to boost the precision. It is however noted that a high precision system is not necessarily important in this module as the intended purpose is mainly to gather and recommend preclinical studies for further research. Model-based systems that could be built in the future may be able to rectify these mistakes and output a higher precision final result.

\subsection{Evaluation of n-grams}
In an earlier iteration of the clinical trial analysis module, only unigrams were used for drug data extraction. This caused an error such as “chloroquine” and “phosphate” being double counted in some instances. Another error included instances where the keyword “interferon” was present in the clinical trial, but the drug dictionary did not have a unigram instance of this keyword. Therefore, the drug was not able to be matched with the dictionary. However, upon adding bigrams and trigrams to this module, “interferon beta” and “interferon alpha” were both successfully extracted from the clinical trial data. 

Likewise, this addition was utilized in the preclinical workflow, but during preliminary analysis of the results, this workflow was ultimately not included in the current module because the initial results showed no significant improvement of relevant data extraction precision for this workflow. The opposite occured and only unfiltered noise data was extracted using n-grams. This is likely due to the fact that unlike the somewhat cleaned and structured clinical trial text, the abstract text is completely freeform, so the backend algorithm best captures different pieces of information such as an experimental value or a reported experiment using single token keywords. The inclusion of multi-word sequences in this workflow can be extremely superfluous and confusing. 

Similarly, n-grams up to trigrams were built into the topic model data preprocessing pipeline. This was because there are some word pairs that are necessary for topics to be accurate and differentiable. One key example shown in Table \ref{tab:tmtable1} is 'sars cov' being one token instead of the two tokens 'sars' and 'cov'. This will give the LDA model cleaner input data especially if differentiating between “SARS” virus and “SARS-CoV-2”.

\subsection{Topic Model Recommendation System}
In Figure 5, the optimal topic model by maximum coherence contained 30 topics. Several topics did not include many papers, so they were excluded in the final results. This filter was done using a data-driven technique where topics that contained less than a fifth of the average amount of papers per topic were excluded in the final results. This meant that topics clustered with very few papers were excluded from the final module results, and the quality of the recommended papers per topic remains clear and differentiable from the evidence shown in Figure 4. This figure showed that most of the Top 5 highest-weighted papers from the corpus in each topic contained something in the title pertinent to this topic. One example is that the hand labeled "AI" topic indeed contained papers with titles discussing deep learning or CT images. Another interesting example is the "PPE" topic where the papers clustered into this topic contained titles talking about N95 respirators, masks, and respiratory aerosols. 
 
 \begin{figure}
	\centering
	\includegraphics[scale=.65]{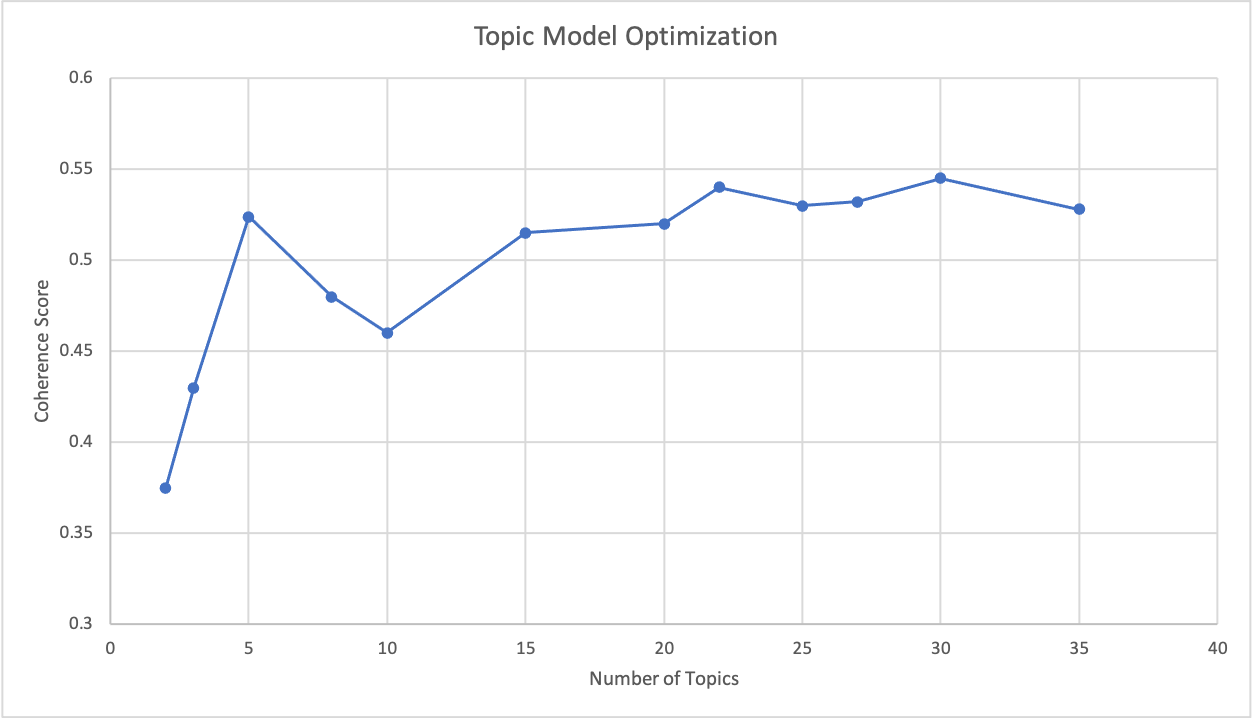}
	\caption{The coherence score compared to number of topics. This gave an optimized number of topics for the final topic model.}
	\label{fig:Picture31}
\end{figure}

\subsection{Limitations and Future Work}
The major limitation for the clinical trial and preclinical text mining section is that they both rely on known drug dictionaries for the exact text search. This means that all drug names contain known, approved, or experimental drugs. However, some experimental drugs have not yet been added into the database, so tracking newly published data on experimental drugs is a major challenge for future AI models. 

The granularity of the preliminary database search was impacted by two important factors: computational cost and precision. In the preclinical text mining and information extraction workflow, we have tried to optimize the performance of the initial preclinical abstract search to have high precision while minimizing the computational cost of text mining. Therefore, during the development of this workflow, we have looked at many different search queries to get the optimal number of abstracts to text mine. Some keywords that were used include: “covid”, “coronavirus”, “preclinical”, “in vitro”, “EC50”, and “experiment”. While all of these keywords yielded the papers of interest, we optimized our search by first using keywords such as “coronavirus” and “sars-cov-2” and then later searched the keywords “EC50” and “IC50”. Compared to searching keywords such as “preclinical” and/or “in vitro”, this allowed a more optimal precision to mining texts of interest while minimizing the number of papers that did not have any useful information as searching the more general keywords gives more samples with useless information and increases the computational effort. 

In our workflow, we primarily used a rule-based text-mining system to extract drug names and experimental values evidenced in Table 1, Table 2, and Figure 3. Due to the nature of the pandemic, the first efforts all revolved around drug repurposing; therefore, the use of a dictionary for text mining should be sufficient for this purpose as all known drug names and aliases were captured in our dictionary. However, if this system were to be tracking a long-term disease or pandemic, a more robust data extraction system should be built. In our preliminary analysis, a Scispacy Named-entity recognition (NER) model\cite{scispacy} was assessed and compared to the rule-based system. Comparing this model in Figure 6, it can be observed that although unique chemicals are identified, the precision is very low compared to a rule-based system especially since it captures information such as "h=15.26" and "ptdtytsvylgkfrg" as chemicals. Future work can include building an NER model with curated data labels from a set of papers in one specific disease area, or an NER model trained with more specific data labels to avoid false positives with experimental values and protein sequences. 

\begin{figure}
	\centering
	\includegraphics[scale=0.5]{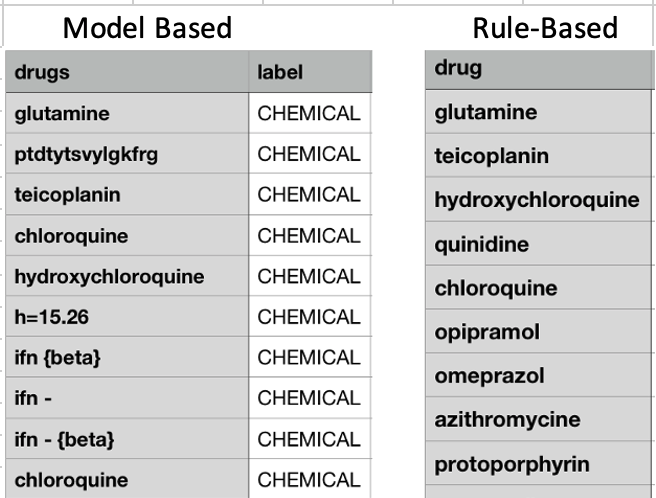}
	\caption{A comparison of an NER model (left) compared to our rule-based system (right).}
	\label{fig:Picture3}
\end{figure}

As an extension to these text mining modules, there are several applications to AI models. Once the data mining modules are mature, data curation efforts for small molecule drugs can be reduced and automated since robust research datasets can be produced\cite{medextractr}. Furthermore, the backend scripts can all be extended and developed for other use-cases. One such use-case can be a module performing data mining on known genes and knock-in, knock-down, or knock-out relationships. Additionally, for the topic model, this can be useful for a variety of scientific fields including cancer or infectious diseases. Future work can look into the topic models of these fields or at a specific area in one of these fields.

\section{Conclusion}
\label{sec:conclusion}
The modules presented on our portal and in this article showcase NLP techniques that may be useful in a global pandemic where lots of text data is being generated daily. Our modules aim to ease the burden of reading thousands of articles daily to the recommended ones automatically updated daily by our text mining systems. This will allow a significant reduction of time spent on reader articles and more time dedicated to research on coronavirus. These modules also have the potential to be scaled to other applications in life sciences and may have use-cases in these areas.

\bibliographystyle{unsrtnat}
\bibliography{references}  

\begin{thebibliography}{15}
\providecommand{\natexlab}[1]{#1}
\providecommand{\url}[1]{\texttt{#1}}
\expandafter\ifx\csname urlstyle\endcsname\relax
  \providecommand{\doi}[1]{doi: #1}\else
  \providecommand{\doi}{doi: \begingroup \urlstyle{rm}\Url}\fi

\bibitem[cov()]{cov19wiki}
Covid-19 pandemic - wikipedia.
\newblock \url{https://en.wikipedia.org/wiki/COVID-19_pandemic}.
\newblock Accessed: 2021-01-30.

\bibitem[tar()]{targetingcov19}
Ghddi targeting covid-19 portal.
\newblock \url{https://ghddi-ailab.github.io/Targeting2019-nCoV/}.
\newblock Accessed: 2021-02-03.

\bibitem[Resources(2020)]{dimensions2020}
Dimensions Resources.
\newblock Dimensions covid-19 publications, datasets and clinical trials, Mar
  2020.
\newblock URL
  \url{https://dimensions.figshare.com/articles/dataset/Dimensions_COVID-19_publications_datasets_and_clinical_trials/11961063/37}.

\bibitem[dru()]{drugbank}
Drugbank.
\newblock \url{https://www.drugbank.ca/}.
\newblock Accessed: 2021-02-03.

\bibitem[fda()]{fdalist}
Drugs@fda: Fda-approved drugs.
\newblock \url{https://www.accessdata.fda.gov/scripts/cder/daf/}.
\newblock Accessed: 2021-02-03.

\bibitem[Davies et~al.(2015)Davies, Nowotka, Papadatos, Dedman, Gaulton,
  Atkinson, Bellis, and Overington]{chembl2015}
Mark Davies, Michał Nowotka, George Papadatos, Nathan Dedman, Anna Gaulton,
  Francis Atkinson, Louisa Bellis, and John~P. Overington.
\newblock {ChEMBL web services: streamlining access to drug discovery data and
  utilities}.
\newblock \emph{Nucleic Acids Research}, 43\penalty0 (W1):\penalty0 W612--W620,
  04 2015.
\newblock ISSN 0305-1048.
\newblock \doi{10.1093/nar/gkv352}.
\newblock URL \url{https://doi.org/10.1093/nar/gkv352}.

\bibitem[kag()]{kaggle2020}
Kaggle -drug treatment extraction (taskvt).
\newblock
  \url{https://www.kaggle.com/benjpjones/drug-treatment-extraction-taskvt/}.
\newblock Accessed: 2021-02-03.

\bibitem[Xu et~al.(2010)Xu, Stenner, Doan, Johnson, Waitman, and Denny]{medex}
Hua Xu, Shane~P Stenner, Son Doan, Kevin~B Johnson, Lemuel~R Waitman, and
  Joshua~C Denny.
\newblock {MedEx: a medication information extraction system for clinical
  narratives}.
\newblock \emph{Journal of the American Medical Informatics Association},
  17\penalty0 (1):\penalty0 19--24, 01 2010.
\newblock ISSN 1067-5027.
\newblock \doi{10.1197/jamia.M3378}.
\newblock URL \url{https://doi.org/10.1197/jamia.M3378}.

\bibitem[Ting(2010)]{ting2010}
Kai~Ming Ting.
\newblock \emph{Precision and Recall}, pages 781--781.
\newblock Springer US, Boston, MA, 2010.
\newblock ISBN 978-0-387-30164-8.
\newblock \doi{10.1007/978-0-387-30164-8_652}.
\newblock URL \url{https://doi.org/10.1007/978-0-387-30164-8_652}.

\bibitem[{\v R}eh{\r u}{\v r}ek and Sojka(2010)]{lda2020}
Radim {\v R}eh{\r u}{\v r}ek and Petr Sojka.
\newblock {Software Framework for Topic Modelling with Large Corpora}.
\newblock In \emph{{Proceedings of the LREC 2010 Workshop on New Challenges for
  NLP Frameworks}}, pages 45--50, Valletta, Malta, May 2010. ELRA.
\newblock \url{http://is.muni.cz/publication/884893/en}.

\bibitem[Shrimp et~al.(2020)Shrimp, Kales, Sanderson, Simeonov, Shen, and
  Hall]{naf1}
Jonathan~H. Shrimp, Stephen~C. Kales, Philip~E. Sanderson, Anton Simeonov, Min
  Shen, and Matthew~D. Hall.
\newblock An enzymatic tmprss2 assay for assessment of clinical candidates and
  discovery of inhibitors as potential treatment of covid-19.
\newblock \emph{bioRxiv}, 2020.
\newblock \doi{10.1101/2020.06.23.167544}.
\newblock URL
  \url{https://www.biorxiv.org/content/early/2020/08/06/2020.06.23.167544}.

\bibitem[Ko et~al.(2020)Ko, Jeon, Ryu, and Kim]{naf2}
Meehyun Ko, Sangeun Jeon, Wang-Shick Ryu, and Seungtaek Kim.
\newblock Comparative analysis of antiviral efficacy of fda-approved drugs
  against sars-cov-2 in human lung cells: Nafamostat is the most potent
  antiviral drug candidate.
\newblock \emph{bioRxiv}, 2020.
\newblock \doi{10.1101/2020.05.12.090035}.
\newblock URL
  \url{https://www.biorxiv.org/content/early/2020/05/12/2020.05.12.090035}.

\bibitem[Yamamoto et~al.(2020)Yamamoto, Kiso, Sakai-Tagawa, Iwatsuki-Horimoto,
  Imai, Takeda, Kinoshita, Ohmagari, Gohda, Semba, Matsuda, Kawaguchi, Kawaoka,
  and Inoue]{naf3}
Mizuki Yamamoto, Maki Kiso, Yuko Sakai-Tagawa, Kiyoko Iwatsuki-Horimoto, Masaki
  Imai, Makoto Takeda, Noriko Kinoshita, Norio Ohmagari, Jin Gohda, Kentaro
  Semba, Zene Matsuda, Yasushi Kawaguchi, Yoshihiro Kawaoka, and Jun-ichiro
  Inoue.
\newblock The anticoagulant nafamostat potently inhibits sars-cov-2 infection
  in vitro: an existing drug with multiple possible therapeutic effects.
\newblock \emph{bioRxiv}, 2020.
\newblock \doi{10.1101/2020.04.22.054981}.
\newblock URL
  \url{https://www.biorxiv.org/content/early/2020/04/23/2020.04.22.054981}.

\bibitem[Neumann et~al.(2019)Neumann, King, Beltagy, and Ammar]{scispacy}
Mark Neumann, Daniel King, Iz~Beltagy, and Waleed Ammar.
\newblock {S}cispa{C}y: {F}ast and {R}obust {M}odels for {B}iomedical {N}atural
  {L}anguage {P}rocessing.
\newblock In \emph{Proceedings of the 18th BioNLP Workshop and Shared Task},
  pages 319--327, Florence, Italy, August 2019. Association for Computational
  Linguistics.
\newblock \doi{10.18653/v1/W19-5034}.
\newblock URL \url{https://www.aclweb.org/anthology/W19-5034}.

\bibitem[Weeks et~al.(2019)Weeks, Beck, McNeer, Bejan, Denny, and
  Choi]{medextractr}
Hannah~L. Weeks, Cole Beck, Elizabeth McNeer, Cosmin~A. Bejan, Joshua~C. Denny,
  and Leena Choi.
\newblock medextractr: A medication extraction algorithm for electronic health
  records using the r programming language.
\newblock \emph{medRxiv}, 2019.
\newblock \doi{10.1101/19007286}.
\newblock URL \url{https://www.medrxiv.org/content/early/2019/09/23/19007286}.

\end{thebibliography}
\end{document}